\documentclass{Interspeech}

\usepackage{multirow}

\interspeechcameraready

\title{Long-Context Speech Synthesis with Context-Aware Memory}

\author[affiliation={1,2}]{Zhipeng}{Li}
\author[affiliation={1},equalcontribution]{Xiaofen}{Xing}
\author[affiliation={1}]{Jingyuan}{Xing}
\author[affiliation={2}]{Hangrui}{Hu}
\author[affiliation={2}]{Heng}{Lu}
\author[affiliation={1,3}]{Xiangmin}{Xu}

\affiliation{}{South China University of Technology}{China}
\affiliation{Speech Lab}{Alibaba Group}{China}
\affiliation{}{Pazhou Lab}{China}
\email{eeleezp@mail.scut.edu.cn, xfxing@scut.edu.cn}

\keywords{text-to-speech, long-context, memory compression}

\usepackage{comment}

\begin{document}

\maketitle

\begin{abstract}
    In long-text speech synthesis, current approaches typically convert text to speech at the sentence-level and concatenate the results to form pseudo-paragraph-level speech. These methods overlook the contextual coherence of paragraphs, leading to reduced naturalness and inconsistencies in style and timbre across the long-form speech. To address these issues, we propose a Context-Aware Memory (CAM)-based long-context Text-to-Speech (TTS) model. The CAM block integrates and retrieves both long-term memory and local context details, enabling dynamic memory updates and transfers within long paragraphs to guide sentence-level speech synthesis. Furthermore, the prefix mask enhances the in-context learning ability by enabling bidirectional attention on prefix tokens while maintaining unidirectional generation. Experimental results demonstrate that the proposed method outperforms baseline and state-of-the-art long-context methods in terms of prosody expressiveness, coherence and context inference cost across paragraph-level speech. Audio samples are available at \href{https://leezp99.github.io/LongContext-CAM-TTS/}{https://leezp99.github.io/LongContext-CAM-TTS/}.
\end{abstract}

\vspace{-1.0em}
\section{Introduction}

In recent years, with advancements in generative models\cite{van2017neural,ho2020denoising,defossez2022high,kumar2024high}, vocoders\cite{kong2020hifi, li2023hiftnet}, and both non-autoregressive\cite{kong23_interspeech,shennaturalspeech,yang24l_interspeech,eskimez2024e2,chen2024f5} and autoregressive models\cite{borsos2023audiolm,wang2023neural,betker2023better}, speech generation technology has reached a level capable of producing natural speech with human-level quality. Benefiting from the high scalability demonstrated by large language models (LLM)\cite{achiam2023gpt,touvron2023llama}, more recent studies\cite{du2024cosyvoice, anastassiou2024seed,lajszczak2024base} have adopted LLM as the core module for text-to-semantic token modeling in TTS tasks, showcasing exceptional natural semantic modeling capabilities.

With the growing demand for applications such as voice assistants, audiobooks, and news broadcasting, the goal of TTS task has gradually shifted from high-quality sentence-level synthesis to coherent and expressive paragraph-level speech. In these long-context scenarios, there exist both explicit and implicit contextual dependencies between historical text and speech. However, current mainstream methods typically split paragraph-level text into sentence-level text and synthesize sentence-level speech individually. This approach neglects the contextual correlation both within and across paragraphs, leading to the following issues: 1) diminished prosodic expressiveness; 2) poor consistency in style, timbre, and speech rate, especially the speech coherence, which severely impacts listeners experience.

Some long-context modeling methods have been proposed recently. Xin et al.\cite{xin2023improving} utilized preceding speech(1 sentence) and bidirectional text context(2–3 sentences) to improve speech prosody; Xiao et al.\cite{xiao23_interspeech} proposed a memory-cached recurrence mechanism based on a fixed-length preceding speech, along with a contextual text encoder;  Xue et al.\cite{xue24c_interspeech} proposed a multi-modal context-enhanced Q-Former that compresses preceding text and speech(5 sentences) to leverage longer contextual information;  Xue et al.\cite{xue24b_interspeech} proposed utilizing CA-CLAP to enhance generation through context retrieval, selecting the whole speech-text prompts (1-2 sentences) as prefix tokens to guide speech generation. While these methods offer valuable insights for long-context TTS, there are still several parts that require further improvement. 1) Speech prosody is text-dependent, so there is a need for a precise context retrieval mechanism for sentence-level speech synthesis; 2) Excessively long guidance prompts will lead to model instability, requiring more concise guidance prompts; 3) Parts of the historical context will be repeatedly used during each inference, which will result in high computational cost; 4) Unable to effectively utilize distant contextual information.  

\begin{figure}[h]
  \setlength{\abovecaptionskip}{0pt}
  \setlength{\belowcaptionskip}{0pt}
  \centering
  \hspace*{-0.1in}
  \includegraphics[width=1.1\linewidth]{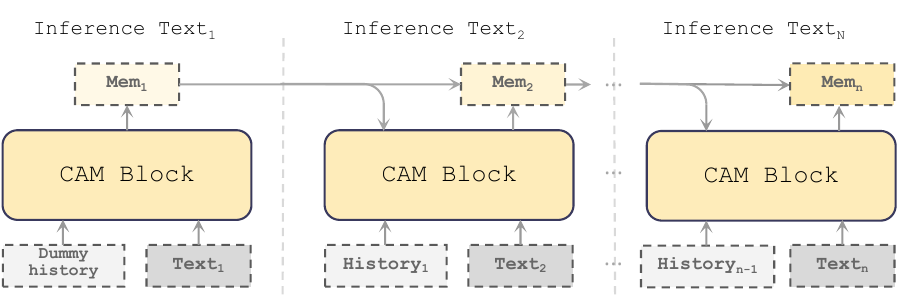}
  \hspace*{-0.1in}
  \caption{Paragraph inference process for CAM block}
\end{figure}

\begin{figure*}[h]
    \setlength{\abovecaptionskip}{0pt}  
    \setlength{\belowcaptionskip}{0pt}  
    \centering
    \hspace*{-0.5in}
    \includegraphics[width=1.12\textwidth]{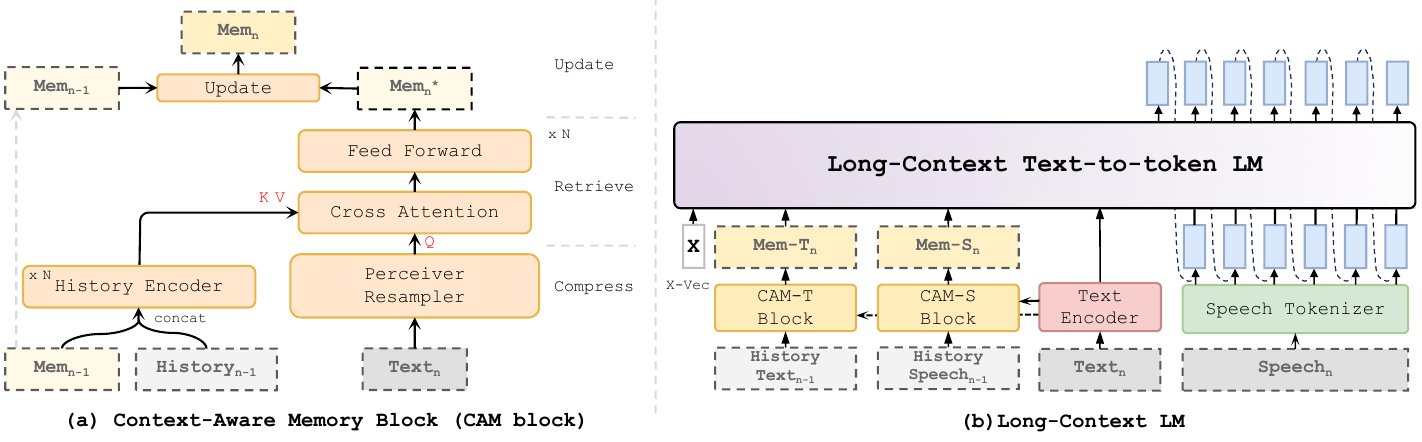}  
    \hspace*{-0.5in}
    \caption{(a) Shows an overview of CAM block, containing compress, retrieve and update three stages. (b) gives an illustration of the Long-Context LM.}
    \label{figure2}
\end{figure*}

Inspired by the infinite attention mechanism with long-term compressed memory proposed by the Google research team\cite{munkhdalai2024leave}, we propose a Context-Aware Memory(CAM) block (Figure 1). The CAM block utilizes perceiver resampler to compress the target text, and separately retrieve the key dependency information from both long-term memory and local context details of historical text and speech. It dynamically updates the memory to guide current speech synthesis and transfers it to subsequent sentences of varying lengths. We integrated the CAM block with the Large Language Model (LLM) to construct a module for long-context text-to-semantic modeling.  To further enhance in-context learning capability, we replace the traditional causal mask with prefix mask, allowing the memory input and text input to freely attend to each other. Compared to methods that incorporate several historical sentences, our solution demonstrates both efficiency and innovation, requiring only a fixed-length long-term memory and the previous one context, which can broaden the model horizon from sentence to paragraph. We summarizes our contributions as follows:

\begin{itemize}
\item We propose a Context-Aware Memory-based long-context TTS model that retrieves and updates memory from both long-term memory and local details to guide high-quality sentence-level speech synthesis within paragraph.
\item We introduce prefix mask to replace the causal mask in LLM, enhancing understanding and in-context learning abilities.  
\item Extensive objective and subjective evaluations show that our proposed method outperforms both baseline and SOTA long-context TTS methods in terms of naturalness, coherence, and inference cost.  
\end{itemize}

\section{Methodology}

The general architecture of Long Context LM model based on Context-Aware Memory block is illustrated in Figure 2. The model consists of two core components: a CAM block for maintaining contextual memory through compression, retrieval, and updates, and a large language model (LLM) for text-to-semantic modeling. Supposing the index of target utterance for synthesis is $n$, we separately use the $(n\text{-}1)$th speech/text as History$\text{-}$Speech$_{n\text{-}1}$/Text$_{n\text{-}1}$. The memory passed down from the $(n\text{-}1)$th speech synthesis is denoted as Mem$_{n\text{-}1}$. These, along with Text$_n$, are used to guide the generation of Speech$_n$.

\subsection{Context-Aware Memory Block}

As shown in Figure 2(a), our proposed CAM block consists of three stages: compression, retrieval, and update. Due to the inherent modal differences between speech and text, we have designed dedicated CAM-Speech and CAM-Text blocks (as CAM$\text{-}$S, CAM$\text{-}$T), which share the same structure but have independent weights. Similarly, memory is divided into Speech Memory (Mem$\text{-}$S) and Text Memory (Mem$\text{-}$T). For simplicity, modal annotations ($\text{-}$S/$\text{-}$T) are omitted in this section. Dummy History in Figure 1 is composed of silence speech and blank text, respectively, designed for the first utterance synthesis where no prior context is available.
\[
Mem_n=CAM(Text_n, Mem_{n\text{-}1}, History_{n\text{-}1})
\]

\indent\textbf{Compress.} We use Perceiver Resampler\cite{alayrac2022flamingo, casanova24_interspeech} to perform cross-attention between variable-length Text$_n$ latent representation and a fixed-length learnable latent query vector. The output of Perceiver Resampler is a compressed fixed-length latent of the target text. The resampling approach enables the model to extract the critical information from the original features. Then, the salient compressed text latent is sent as a query to the Cross-Attention (in Retrieve stage).

\indent\textbf{Retrieve.} Since the long-term memory Mem$_{n\text{-}1}$ is retrieved from the previous utterance Text${_{n\text{-}1}}$ and does not contain History${_{n\text{-}1}}$, it needs to be fused first. We concatenate and feed them into a Transformer-based History Encoder. Then, using the target text representation obtained from the compression stage (as Query), multiple retrievals are performed on the fused contextual information (as Key and Value) to capture the most critical contextual dependencies Mem$_{n}^*$ in the current utterance.

\indent\textbf{Update.} After retrieval, we update the memory and obtain next states. We aggregate the long-term memory Mem$_{n\text{-}1}$ and memory  retrieved value Mem$_{n}^*$ via a learned scalar \(\alpha\), allowing a dynamic trade-off between long-term and local context.
\[
Mem_n=sigmoid(\alpha)\odot Mem_{n}^{*}+(1-sigmoid(\alpha))\odot Mem_{n\text{-}1}
\]
After compression, retrieval, and update, the latest memory representations, Mem\text{-}S${_n}$ and Mem\text{-}T${_n}$, are obtained, which integrate both long-term context and local content. These representations are then fed into the Long-Context LM to guide the text-to-semantic modeling. 

\subsection{Long Context LM}

To enhance prosody expression and coherence, we use the contextual memory Mem${\text{-}T_n}$ and Mem${\text{-}S_n}$ generated by the above modules as inputs to the LLM. The main architectural follows CosyVoice\cite{du2024cosyvoice}, we use X-vectors from Cam++\cite{wang2023cam++}, and employ paragraph-level X-vectors rather than utterance-level during both the training and inference phases. This allows LLM to have larger learning space, thereby enhancing the naturalness and coherence of generate speech. The LLM input is as follows:
\[
[X_{vec},Mem\text{-}T_n,Mem\text{-}S_n,TE(Text_n),ST(Speech_n)]
\]
TE and ST are Text Encoder and Speech Tokenizer, respectively. The pre-trained Flow Matching and Vocoder are applied to convert the generated tokens into waveforms.

In Long-Context LM, X$_{vec}$,Mem\text{-}T$_n$,Mem\text{-}S$_n$ are treated as pre-filled information. Therefore, during training, only the cross-entropy loss of the generated speech tokens are considered.

\subsection{Prefix Mask For LLM}

Nowadays, most text-to-semantic LLMs in TTS belong to a decoder-only architecture. These models typically use causal mask, where each token can only attend to preceding tokens and itself. However, applying    
causal mask strictly across the whole sequence during training may limit the model's performance\cite{ding2023causallm}, especially for Long-Context LM with in-context learning capabilities. Therefore, we introduce the Prefix Mask (Figure 3) for Long-Context LM, which applies bidirectional attention to the prefix tokens, such as X-vec, Mem, and Text, allowing bidirectional encoding of prefix sequences along the temporal dimension. Unidirectional attention is maintained on the generated tokens to ensure the coherence of the  generation.

\begin{figure}[h]
  \setlength{\abovecaptionskip}{-6pt}
  \centering
  \hspace*{-0.3in}
  \includegraphics[width=1.1\linewidth]{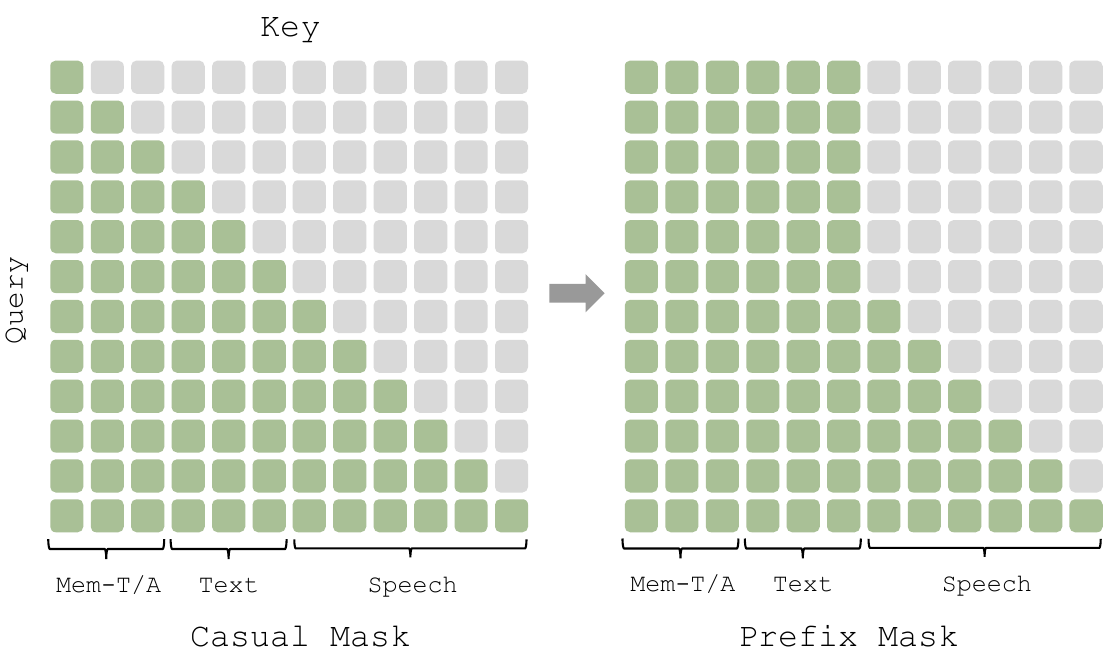}
  \hspace*{-0.3in}
  \caption{Illustration of Casual Mask and Prefix Mask in LLM}
  \label{figure1.}
\end{figure}

\section{Experiments}
\subsection{Datasets}
To train our proposed Long-Context LM, we collected about 15,000+ hours of Chinese mandarin audiobooks from Internet, including about 75000+ complete novel chapters. The data primarily consists of single-podcast speech. We utilize the Demucs\cite{rouard2023hybrid} to extract clean human vocals from raw speech data. We split the long segments into smaller ones, each utterance is under 30 seconds, and the average duration of 16.2 seconds. Paraformer\cite{gao22b_interspeech} is used to transcribe the data. We randomly selected 100 additional complete chapters for validation.

\subsection{Experimental Setup}
\textbf{Training.} LongContext LM is modified from CosyVoice-LM, SpeechTokenizer is 50Hz version. For Long-Context LM, we train from scratch with a constant learning rate of $10^{-4}$. In CAM Block, the Perceiver Resampler produces a fixed number of 32 embeddings. The History Encoder consists of two stacked Transformer blocks and employs two layers of Retrieve stages. All the models were trained for 100M steps with a dynamic batch size of 10,000 tokens per batch to ensure complete  convergence.

\noindent\textbf{Inference.} Random sampling decoding strategy is employed for all LM models.

\subsection{Model Evaluation}
\subsubsection{Compared Methods}
To evaluate the performance of our method, we compare it with state-of-the-art Long-Context TTS systems.

\noindent$\bullet$ \textbf{MMCE\text{-}Qformer} Xue et al.\cite{xue24c_interspeech} proposed a multi-modal context-enhanced Qformer, utilizing compressive long-context information to improve TTS performance.

\noindent$\bullet$ \textbf{CLAP\text{-}RAG} Xue et al.\cite{xue24b_interspeech} proposed a RAG-enhanced prompt-based TTS framework using a context-aware contrastive language-audio pretraining model. And it utilizes entire prompts to guide the generation process.

\noindent$\bullet$ \textbf{Proposed} Our proposed Long-Context LM with context-aware memory block and prefix mask.

We reproduced the MMCE\text{-}Qformer and CLAP\text{-}RAG models on the Cosyvoice\text{-}LM backbone following the original paper's implementation, with the context lengths set to 5 and 1.

\subsubsection{Ablation Study}

We perform ablation studies to evaluate the effectiveness of key modules in Long-Context LM. 

\noindent$\bullet$ \textbf{Baseline} LM trained from scratch using Cosyvoice-LM\cite{du2024cosyvoice} as backbone.

\noindent$\bullet$ \textbf{w/o Mem\text{-}T} Long-Context LM without CAM\text{-}T block.

\noindent$\bullet$ \textbf{w/o Mem\text{-}S} Long-Context LM without CAM\text{-}S block.

\noindent$\bullet$ \textbf{w/o prefix mask} Long-Context LM with standard  casual mask.

\subsubsection{Evaluation Metrics}
We use both objective and subjective metrics to evaluate the aforementioned models.

\noindent\textbf{Subjective Evaluation} We randomly choose 20 sentence-level (about 15s) speech samples and 10 long-form (about 60s) speech samples for evaluation. long-form speech is constructed by combining multiple sentence-level speech. We conduct paragraph MOS (mean opinion score) to evaluate the expressiveness and naturalness of the ground truth recording and sentence-level synthetic speech. Paragraph CoMOS (Consistency Mean Opinion Score) is used to evaluate the overall coherency in style and timbre throughout the long-form speech. In each MOS test, 10 native Chinese Mandarin listeners rate the MOS and CMOS on a scale from 1 to 5 with 0.5 point intervals. The final scores are reported with confidence interval of 95\% to ensure statistical reliability.

\noindent\textbf{Objective Evaluation} For the objective metrics, we evaluate speaker similarity (SIM), robustness (CER), and speech quality (SpeechBertScore). Specifically, for speaker similarity, we compute the cosine similarity between the speaker-level X\text{-}vector used in the LM during inference and the X-vector of the generated samples, the mean represents overall similarity, and the variance indicates timbre consistency stability. For robustness, Paraformer-zh is employed as the ASR model to evaluate the content consistency. For speech quality, we use SpeechBERTScore\cite{saeki24_interspeech} for quality estimation, as it shows higher human rating correlation compared to previous methods.

\noindent\textbf{Inference Context Cost} In sentence-level speech synthesis, Num represents the number of sentence-level contexts used by each long-context method,  and Prefix Len refers to the number of context-related tokens fed into the LM model.

\begin{table*}
\vspace{0.5em}
    \centering
    \normalsize
    \setlength{\extrarowheight}{4pt}
    \caption{Evaluation Results of the Proposed Method, SOTA Long-Context Methods, and Ablation Studies.}
    \begin{tabular}{l@{\hspace{5mm}}|c@{\hspace{5mm}}c@{\hspace{5mm}}|c@{\hspace{5mm}}c@{\hspace{5mm}}c@{\hspace{5mm}}|c@{\hspace{4mm}}|c@{\hspace{4mm}}}
        \hline
        \addlinespace[+1.2mm] 
        \hline
        \multirow{2}{*}{} & \multicolumn{2}{c|}{Subjective} & \multicolumn{3}{c|}{Objective} & \multicolumn{2}{c}{Context Cost} \\ \cline{2-8}
         & MOS ($\uparrow$) & CoMOS ($\uparrow$) & SpeechBERT ($\uparrow$) & CER ($\downarrow$) & SIM ($\uparrow$) & Num & Prefix Len \\
        \hline
        Ground Truth & $4.406_{\pm0.095}$ & $4.870_{\pm0.056}$ & $100$ & $4.286\%$ & $93.716_{(0.046)}$ & $-$ & $-$ \\
        \hline
        MMCE\text{-}Qformer & $3.557_{\pm0.082}$ & $3.885_{\pm0.112}$ & $79.031$ & $5.075\%$ & $85.110_{(0.021)}$ & $5$ & Fixed: $64$ \\
        CLAP\text{-}RAG & $3.489_{\pm0.133}$ & $3.717_{\pm0.183}$ & $78.892$ & $6.234\%$ & $84.920_{(0.037)}$ & $1$ & Variable \\
        \hline
        Baseline & $3.468_{\pm0.113}$ & $3.460_{\pm0.120}$ & $77.776$ & $5.850\%$ & $85.051_{(0.035)}$ & $-$ & $-$ \\
        \hline
        \textbf{Proposed} & $\textbf{3.796}_{\pm\textbf{0.091}}$ & $\textbf{3.992}_{\pm\textbf{0.127}}$ & $\textbf{80.448}$ & $\textbf{4.140}\%$ & $\textbf{85.685}_{(\textbf{0.019})}$ & $1$ & Fixed: 64 \\
        {\quad w/o Mem\text{-}T} & $3.604_{\pm0.146}$ & $3.887_{\pm0.135}$ & $78.358$ & $5.036\%$ & $85.461_{(0.031)}$ & $1$ & Fixed: 32 \\
        {\quad w/o Mem\text{-}S} & $3.516_{\pm0.155}$ & $3.827_{\pm0.129}$ & $78.063$ & $5.496\%$ & $85.150_{(0.039)}$ & $1$ & Fixed: 32 \\
        {\quad w/o prefix mask} & $3.661_{\pm0.158}$ & $3.846_{\pm0.125}$ & $80.243$ & $4.633\%$ & $85.135_{(0.026)}$ & $1$ & Fixed: 64 \\
        \hline
        \addlinespace[1.2mm] 
        \hline
    \end{tabular}

    \label{tab:my_label}
\vspace{0.5em}
\end{table*}

\vspace{0.5em}
\subsection{Experimental Results}

\subsubsection{Performance comparison}

We conducted a comparison of three long-context methods. First, we analyzed the context costs in inference. For each sentence-level speech synthesis, MMCE\text{-}Qformer takes five contexts as input and generates 64 tokens as prefix tokens to guide the generation; CLAP-RAG retrieves the most relevant sentence from all contexts using CLAP and utilizes the complete text \& speech ($\sim$900 tokens) as length-variable prefix tokens. This method places an immense computational burden on the key-value(KV) cache. In contrast, Proposed Method combines the strengths of both approaches, requiring only the previous context History$_{n\text{-}1}$ and memory Mem$_{n\text{-}1}$ as input, while generating 64 memory tokens for synthesis. This significantly reduces consumption of retrieval and autoregressive inference. 

The subjective evaluations MOS and CoMOS indicate that using the retrieved complete prompt for guiding generation (CLAP\text{-}RAG) improves coherence compared to the baseline, but shows no significant improvement in naturalness. Meanwhile, the Proposed Method outperforms MMCE\text{-}Qformer and CLAP\text{-}RAG in overall performance. In the objective test, the Proposed Method shows better mean and variance in SIM, indicating that it generates speech with the most similar and stable timbre to the target speaker. In SpeechBERTScore, Proposed Method also slightly outperforms the other two models. We attribute this advantage to the introduction of Memory Tokens, which continuously update the memory by balancing latest context with long-term information, thus guiding speech generation through the key contextual cues. Additionally, CLAP\text{-}RAG exhibits the worst performance in CER, which we attribute to the increased hallucination effects during inference caused by excessively long prompts in the generated sequences. In contrast, the Proposed Method employs a fixed number of Prefix Tokens, mitigating inference instability and enhancing robustness (CER).

\subsubsection{Ablation Analysis Results}
We conduct ablation studies to explore the influence of each component in Proposed method.

The experimental results indicate that the baseline model, which lacks the text and speech memory modules, fails to leverage contextual information for guidance, resulting in suboptimal performance in expressiveness and coherence. Moreover, through the MOS and CoMOS tests, we found that the methods with in-context learning capabilities that utilize prefix tokens to guide generation benefit noticeably from the prefix mask. This suggests that the prefix mask improves the model's ability to generate context-driven predictions. Furthermore, we compared the effects of speech memory and text memory, and the results indicate that speech context information yields better performance. We attribute this to the one-to-many relationship between text and speech, where a single textual input can correspond to multiple valid speech outputs with variations in prosody and emotion. Compared to textual context, speech context inherently captures richer acoustic and prosodic information, making speech memory embeddings more effective in enhancing the coherence and timbre consistency of the generated speech. Additionally, through data inspection, we found that the CER scores of ground truth were influenced by the source separation technique Demucs, with some data showing a reduction in vocal quality, resulting in higher CER scores. Despite this challenge, the robustness of the LLM-based model helps to compensate for these degradations, resulting in CER scores that are lower than those of the ground truth data.

Finally, it should be noted that although the proposed method shows promising performance, there remains a noticeable gap in naturalness and coherence compared to real audiobooks data (Ground Truth). There is still significant potential for improvement in paragraph-level speech generation, which warrants further research and refinement in the future.

\vspace{1.0em}
\section{Conclusions}

In this study, we propose an effective long-context TTS model that leverages compressed context-aware memory to enhance both naturalness and coherence in sentence-level speech synthesis. The CAM block integrates and retrieves both long-term memory and local context details, dynamically updating the memory to maintain the key contextual history within paragraph. The latest context memory is used as prefix information to guide token generation in the LM model, with a prefix mask enhancing in-context learning. Experiments on the Chinese mandarin audiobook corpus demonstrate that the proposed method achieves greater expressiveness, coherence, and lower context computation cost in paragraph reading compared to both the baseline model and previous long-context methods.

\section{Acknowledgements}

The work is supported in part by Nansha Key Project under Grant 2022ZD011; in part by Guangdong Basic and Applied Basic Research Foundation (2025A1515011203); in part by Guangdong Provincial Key Laboratory of Human Digital Twin (2022B1212010004).

\bibliographystyle{IEEEtran}
\bibliography{mybib}

\end{document}